# Detection of Student Disengagement in Online Classes Using Deep Learning: A Review

Ahmed Mohamed
*Systems and Biomedical Engineering*
*Cairo University*

Mostafa Ali
*Systems and Biomedical Engineering*
*Cairo University*

Shahd Ahmed
*Systems and Biomedical Engineering*
*Cairo University*

Nouran Hani
*Systems and Biomedical Engineering*
*Cairo University*

Mohammed Hisham
*Systems and Biomedical Engineering*
*Cairo University*

Meram Mahmoud
*Systems and Biomedical Engineering*
*Cairo University*

Student Engagement was first introduced in the educational context in the 1980s, to understand and address issues such as student boredom and dropout rates, as discussed in Newmann's study that conceptualized student engagement. One of the challenges is the lack of students' participation in the classes which is part of learning disengagement.[1]

Axelson and Flick defined Student engagement as: ***"How involved or interested students appear to be in their learning and how connected they are to their classes, their institutions, and each other"***[2]. This concept has been supported by a survey conducted on high school students, which found a direct positive relationship between student engagement and academic performance, as the more the student engaged in the study material, the higher grades they get. [3]

As the definition of engagement has become more complex, it has been divided into various components. One model, according to Fredric Categorization of learners engagement[4], divides engagement into three components: behavioral, cognitive, and emotional. Similarly, Bosch divides engagement into three categories: affective, behavioral, and cognitive.[5] In contrast, Anderson's research expanded the components of engagement to include behavioral, cognitive, academic, and psychological dimensions [6]

In conclusion, based on several readings, learner engagement can be defined as the feelings that drive actions reflecting a student's interest in academic material. These actions include both individual and peer activities, occurring in both classroom settings and extracurricular activities.

Now, Online learning faces different challenges, one of which is the lack of student participation in classes leading to disengage in the session and drop out from the learning process. [7] This lack of engagement was observed during the shifting to online education in response to the COVID-19 pandemic when online learning had become the new norm of education especially after the closure of educational institutions in over 200 countries.[8]

Assessing and measuring student engagement during class is a key factor in solving the problem of disengagement, particularly in online settings. It enables educators to track the attention levels of their students [9]. Methods for measuring student engagement can be categorized into three main types based on the level of learner involvement in the detection process: automatic, semi-automatic, and manual.[10]

In our literature, we are concerned about the manual assessment of student engagement using Artificial Intelligence (AI) technologies, with a focus on Deep Learning models.

We have collected relevant studies after using this search string - **("learners" OR "students") AND ("online learning" OR "online education") AND ("Engagement Assessment" OR "Engagement Tracking" OR "Disengagement Detection") AND ("computer vision" OR "Deep learning")** -which was created and subsequently applied to the Google Scholar database for searching titles and abstracts to find relevant articles.

In order to set our inclusion and exclusion criteria for study selection we have followed Cooper's guidelines, [11] :

*Inclusion criteria:*
- **Language**: Studies must be written in English.
- **Study type:** Accepted study types include empirical studies (full articles, papers, notes, extended abstracts, and work-in-progress papers).
- **Peer-Reviewed:** Studies should have undergone peer review.
- **Focus:** Studies must exclusively focus on Online Learning.
- **Length**: A minimum of four pages is required.
- **Engagement Focus:** Research must explicitly focus on learners' engagement in Online Learning, offering insights into enhancing engagement or deepening the understanding of the topic.
- **Publication date:** Studies published between January 1, 2013, and November 1, 2023, are eligible.

- **No Replication:** Studies should not duplicate the same idea by the same author (s).
- **Source Types:** Both journal articles and papers included in conference proceedings are acceptable.

*Exclusion criteria:*
- **Non-English Language:** Studies not written in English are excluded.
- **Irrelevant Source Types:** Blog posts, magazine articles, theses, newsletters, and literature review articles or papers do not meet the inclusion criteria.
- **Repetition by Same Authors:** Repeated contributions by the same authors in journal articles and conference papers are not eligible.

The search process initially identified 272 studies, but after applying our exclusion and inclusion criteria, only 72 study met our criteria. We have established a CSV[1] spreadsheet to systematically gather excerpts from all these studies, organizing the information in rows and columns. Each row provided a summary of the data extracted from each study, while the columns detailed the types of data being extracted. These columns typically included information such as the title, author, publication year, article or paper type, and other relevant data. Furthermore, we have chosen out of the 72 study only 38 study[2] after reviewing the results once more.

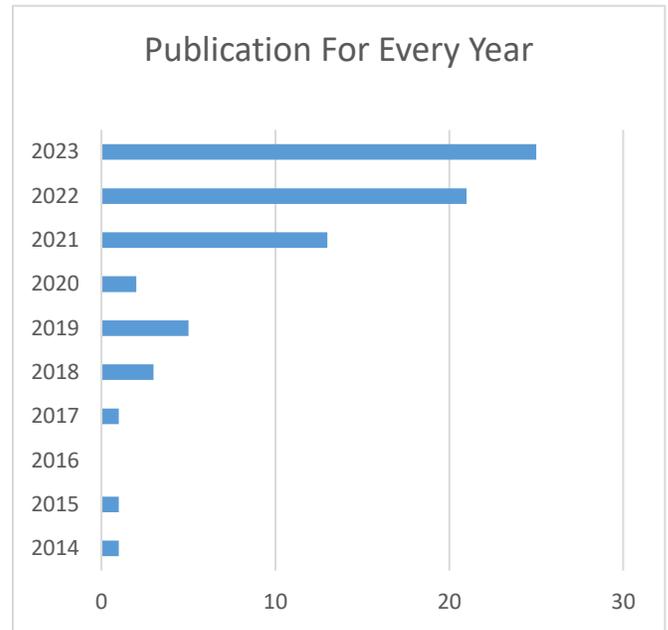

*Figure 2 Numbers of Publications Found Every Year*

In our research, we are concerned about utilizing AI for assessing engagement, particularly focusing on computer vision techniques due to their ease of use and the widespread availability of low-cost cameras. Additionally, we are exploring the use of affective computing techniques, which closely resemble teacher observations and do not disrupt the learning process for the students during the session.

As stated, 38 papers from our results were found that depend on computer vision. These papers presented various approaches for detecting disengagement, along with different methods and models, which will be discussed below.

We examined the collected research to gather answers to our research questions, which are as follows:
1. Which methods or technologies are used to assess students' engagement?
2. Which indicators or specific acts are assessed to detect disengagement?

Additionally, we gathered demographic information about the studies. The studies were published from 2014 to the present date, with the year 2023 having the highest publications count Figure 2. The major focus of these studies was on students, with sample sizes ranging from 19 to 432 students. Their ages vary between 17 to 29 years. A detailed description of every paper sample was added to the studies data sheet mentioned earlier.

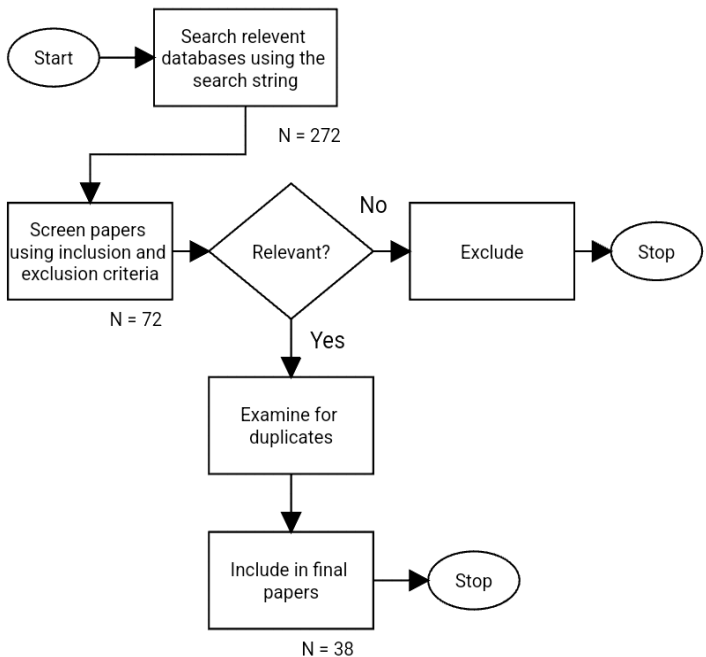

*Figure 1 Studies Selection Process*

---

[1] All the 72 researches with details are listed in this worksheet: Selected Studies details

[2] Detailed description of our 38 main review studies: Main review studies

Computer vision-based models offer various ways to assess learners' engagement, with the most common modalities being facial expressions, gestures, postures, and eye movements. Out of 37 papers investigated, 21 of them focused on facial expressions. Figure [3] Therefore, we can categorize the approaches into two main groups: Face-dependent and Face-independent Figure [4].

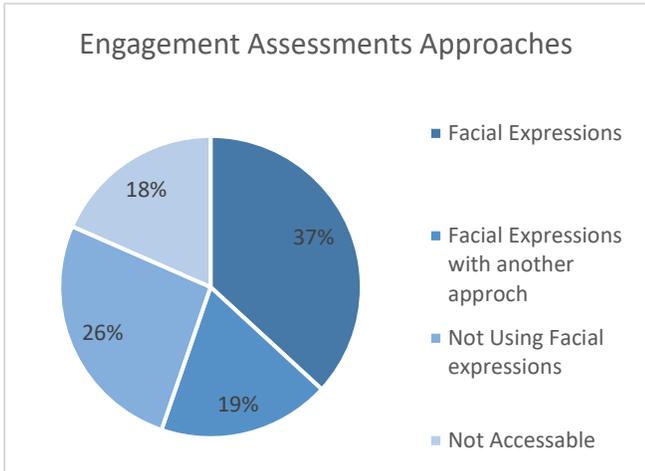

Figure 3 Engagement Assessments Approaches Found in Review Papers

Face-dependent assessments can be further divided into two types: part-based indicators and appearance-based indicators. Part-based indicators concentrate on specific facial features such as eyes and mouth. The Facial Action Coding System (FACS) is utilized to analyze different facial parts, referred to as Action Units (AUs), which represent the movements of facial muscles. These units are used to measure specific emotions [12]. These emotions then reflect the level of student engagement with the learning material.

In [13] and [14], an open-source toolkit called OpenFace 2.0 is employed for various tasks, including detecting face landmarks, recognizing a set of Action Units (AUs), tracking eye gaze, and positioning the head. Using this technology, 30 features are extracted from video frames of students. Subsequently, a Long Short-Term Memory (LSTM) deep neural network is used to process the data and detect engagement in a time series model.

Appearance-based models, on the other hand, rely on extracting features from the entire face and generating patterns for engagement classification. In [15], a lightweight attentional convolutional neural network (CNN) is introduced for face expression recognition. This model recognizes four main expressions, each of which is assigned a weight and contributes to the assessment of engagement based on specific thresholds and equations, dependent on a trained CNN model.

In our literature, approximately 10 studies have focused on indicators that are not directly related to facial expressions but rather on user activities during the session, such as mouse activity or user posture. In [16] and [17], learner's mouse activity is integrated with their gaze during the session and used to train AI models for detecting learner disengagement in online settings. The first study employed a CNN model trained to establish a connection between mouse activity and gaze, while the other study used data from a trained Support Vector Machine (SVM) model to analyze user mouse activity in a learning session to detect disengagement. They achieved this using their dataset of computer users' mouse activities during specific tasks.

In another study [18], the activity data of 360 students within the e-learning platform was collected and summarized into eight features, which included Total Logins, Activity inside the content area, Number of Clicks, Join Session, User Activity Group, Time Spent, Total Items, and Time Spent in Session Attendance. Subsequently, this labeled data was used as input for two models: an SVM model and an Artificial Neural Network.

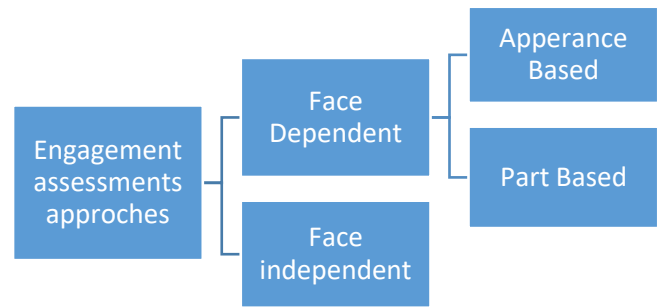

Figure 4 Engagement Assessment Approaches

Based on our review, we have not found yet any studies that specifically illustrate the level of engagement of learners based on their drowsiness status. As mentioned earlier [10], drowsiness is considered one of the most significant indicators of disengagement during online learning. Using various deep learning models, our goal is to introduce an innovative framework for assessing student engagement by considering both facial expressions and drowsiness status. This model is proposed as a web app solution for instructors and education providers, which will assist them in continuously monitoring their learners' levels of engagement.